# Inferring Company Structure from Limited Available Information

**Mugurel Ionut Andreica, Politehnica University of Bucharest, mugurel.andreica@cs.pub.ro**
**Romulus Andreica, Angela Andreica, Commercial Academy Satu Mare**

**Abstract:** *In this paper we present several algorithmic techniques for inferring the structure of a company when only a limited amount of information is available. We consider problems with two types of inputs: the number of pairs of employees with a given property and restricted information about the hierarchical structure of the company. We provide dynamic programming and greedy algorithms for these problems.*

## 1 Introduction

There are many situations in which an in-depth analysis of a company needs to be performed, but only limited information is available about its structure and hierarchy. In this paper we present algorithmic techniques for inferring the structure of a company based on very limited available information. We consider two types of problems, based on the input which is provided: problems where the input contains the number of pairs of employees with a given property and problems where the input contains restricted information about the hierarchical structure of the company. We provide dynamic programming and greedy algorithms for these problems. The rest of this paper is structured as follows. In Sections 2, 3 and 4 we discuss problems with the first type of input. In Section 5 we discuss a problem having the second type of input. In Section 6 we present related work and in Section 7 we conclude.

## 2 Partitioning the Employees into Departments

A company has n employees and d departments. We know that there are k pairs of employees working in the same department. We want to infer the values $e_1$, $e_2$, ..., $e_d$, where $e_i$ is the number of employees working in department i ($0 \leq e_i \leq n$). The following condition must hold

$$k = \sum_{i=1}^{d} C_{e_i}^{2} \qquad (1)$$

Moreover, we must have $k \leq n \cdot (n-1)/2$. We will solve a slightly different problem first: given n and k, determine the minimum number of departments d the company may have. We provide a dynamic programming algorithm for this problem. We will compute a table $D_{min}(i,j)$ representing the minimum number of departments the company must have if there are i employees and j pairs of employees working in the same department. We have $D_{min}(0,0)=0$ and $D_{min}(0,j>0)=+\infty$. For i>0 and j=0 we have $D_{min}(i,j)=i$; for i>0 and $0<j\leq i\cdot(i-1)/2$ we have:

$$D_{min}(i, j) = 1 + \min_{1 \leq p \leq i} \{D_{min}(i - p, j - C_{p}^{2})\} \qquad (2)$$

The value of p represents the number of employees in the first department (with i-p employees and j-p·(p-1)/2 pairs of employees remaining for the other departments).

For j>i·(i-1)/2, we have $D_{min}(i,j)=+\infty$. Since the values j>k are of no interest to us, it is easy to see that the time complexity of the algorithm is $O(n^2 \cdot k)$. At a first glance, this problem can be solved by a simple greedy algorithm:

**SimpleGreedyAlgorithm(n,k):**
*d=0*
**while** (*k>0*) **do**
 *find the largest p such that p·(p-1)/2≤k*
 *k=k-p·(p-1)/2*
 *n=n-p*
**return** d+n

However, a counterexample is easy to give (e.g. for n=12 and k=18). Now we can solve the initial problem. If $D_{min}(n,k) \leq d$, then we have a solution and we can easily compute this solution by tracing back the way the values in the table $D_{min}$ were computed (we store the value of p which minimized the value of each entry (i,j)). The departments $D_{min}(n,k)+1, ..., d$ will contain no employees.

## 3 Partitioning Bosses and Simple Employees into Departments

A company has an unknown number d of departments. Within each department i (1≤i≤d), there is an unknown number of bosses $b_i$ ($b_i \geq 1$) and an unknown number of simple employees $e_i$ ($e_i \geq 1$). Each boss interacts with each simple employee in his department. The only information available is the total number of interactions TI:

$$TI = \sum_{i=1}^{d} b_i \cdot e_i \qquad (3)$$

For this problem, a trivial valid solution consists of 1 department, 1 boss and TI simple employees. Therefore we will be interested in finding the structure with the minimum total number of employees (bosses plus simple employees) and, in case of ties, that with a minimum number of bosses. In order to infer an optimal company structure (the number of departments and the number of bosses and simple employees within each department), we will use a dynamic programming algorithm. We will compute the tables $TE_{min}(i)$=the minimum total number of employees if the total number of interactions is i and $B_{min}(i)$=the minimum number of bosses if the total number of interactions is i and the total number of employees is $TE_{min}(i)$. We have $TE_{min}(0)=B_{min}(0)=0$. The pseudocode of the algorithm is given below:

**BossesAndEmployeesDynamicProgramming(TI):**
$TE_{min}(0)=B_{min}(0)=0$
**for** *i=1* **to** *TI* **do**
　$TE_{min}(i)=B_{min}(i)=+\infty$
　**for** *b=1* **to** *i* **do**
　　**for** *e=1* **to** *i/b* **do**
　　　**if** (($TE_{min}(i-b \cdot e)+b+e<TE_{min}(i)$) **or**
　　　(($TE_{min}(i-b \cdot e)+b+e=TE_{min}(i)$) **and** ($B_{min}(i-b \cdot e)+b<B_{min}(i)$))) **then**
　　　　$TE_{min}(i)=TE_{min}(i-b \cdot e)+b+e$
　　　　$B_{min}(i)=B_{min}(i-b \cdot e)+b$

The solution to our problem can be constructed by tracing back the way the entries in the tables $TE_{min}$ and $B_{min}$ were computed. The time complexity of the algorithm is $O(1/1+2/1+2/2+...+i/1+i/2+...+i/i+...+TI/1+TI/2+...+TI/TI)=O(TI^2 \cdot ln(TI))$.

## 4 Inferring the Communication Structure from Critical Pairs

A company has n employees and some pairs of them have the ability (permission) to communicate directly.

Every employee can communicate (directly or indirectly, through intermediate employees) with all the other employees. The only pieces of information available are n and k, where k denotes the number of critical pairs of employees. A pair of employees (i,j) is critical if there exists a pair of employees (p,q) which can communicate directly, but if their direct communication ability ceased, then the employees i and j would not be able to communicate at all (directly or indirectly). It is possible that one or both employees of the pair (p,q) belong to the pair (i,j). We will describe the communication structure of the company as an undirected graph, where each vertex corresponds to an employee and each edge corresponds to a pair of employees who can communicate directly. Thus, if all the paths between a pair of vertices (i,j) pass through a single edge (p,q), then the pair (i,j) is critical. We are interested in inferring the communication graph of the company. We must first notice that if a pair (p,q) is a critical edge [1] in the communication graph and there $n_1$ employees on one side of the edge and $n_2$ employees on the other ($n_1+n_2=n$), then the structure contains $n_1 \cdot n_2$ critical pairs plus the number of critical pairs in the part of the graph with $n_1$ employees and the number of critical pairs in the part of the graph with $n_2$ employees. Let's consider the tree of connected components of the communication graph, where each connected component of the graph is a tree vertex and each critical edge of the graph is a tree edge. Let's consider a leaf vertex in this tree. We can reattach the leaf vertex to any other tree vertex (thus changing the underlying communication graph) and maintain the same number of critical pairs. Thus, we could modify the tree of connected components into a path and still have the same number of critical pairs. With this observation, we will use a dynamic programming algorithm and compute the following table: *OK(i,j)*=true, if it is possible to build a communication structure consisting of i employees and containing j critical pairs (and false, otherwise). We have OK(0,0)=OK(1,0)=OK(i≥3,0)=true (if i≥3, then we can build a clique or a cycle with the i vertices of the communication graph), but OK(2,0)=false. We also have OK(i≥0, j>i·(i-1)/2)=false. For all the other situations:

$$OK(i, j) = \begin{cases} \text{true, if } \exists (p \in \{1,3,4,\ldots,i\}).OK(i-p, j-p\cdot(i-p)) = \text{true} \\ \text{false, otherwise} \end{cases} \quad (4)$$

If *OK(n,k)=true*, we can construct the communication graph by tracing back the computation of the *OK* table entries. The time complexity of the algorithm is $O(n^2 \cdot k)$.

## 5 Inferring the Hierarchical Structure of a Company

A company has n employees organized into a hierarchical structure (each employee has one boss, except the company manager). We will consider this structure a rooted, directed tree, where each vertex corresponds to an employee and the parent of a vertex corresponds to the employee's boss. Furthermore, the vertices are numbered from 1 to n. We only know two orderings of the tree vertices: the ordering corresponding to the depth-first (DF) traversal of the tree (df(1), df(2), ..., df(n)) and the one corresponding to the breadth-first (BF) traversal (bf(1), bf(2), ..., bf(n)). For both traversals, the first visited vertex is the tree root. The children of a vertex are ordered in ascending order of their numbers and are visited (expanded) according to this order (in both traversals). Based on these two orderings, we want to infer the tree structure of the company. We first compute for each vertex i the values *posdf(i)* and *posbf(i)*, the positions (from 1 to n) of vertex i in the DF and BF orderings (i.e. *df(posdf(i))=i* and *bf(posbf(i))=i*). We now present a linear (O(n)) algorithm solving this problem. We compute a value *parent(i)* for each vertex i. Initially, all the *parent* values are set to 0. At the end of the algorithm, only the root vertex r (obviously, r=df(1)=bf(1)) will have parent(r)=0 and the parent values will define a tree structure which is consistent with the given DF and BF orderings. The main function of the algorithm is the function *Compute*, which has four parameters: *v*, *df_max_pos*, *bf_min_pos* and *bf_max_pos*. The first call will be *Compute(r, n, 2, n)*. We will consider that *df(n+1)=bf(n+1)=0*.

**Compute(v, df_max_pos, bf_min_pos, bf_max_pos):**
**if** ((*posdf(bf(bf_min_pos))≤df_max_pos*) **and**
  (*posdf(bf(bf_min_pos))>posdf(v)*) **and**
  (*parent(bf(bf_min_pos))=0*)) **then**
  *parent(bf(bf_min_pos))=v*
  *last_son=bf(bf_min_pos); pos_last_son=bf_min_pos*
  **for** *i=bf_min_pos+1* **to** *bf_max_pos* **do**
    **if** ((*posdf(bf(i))≤df_max_pos*) **and** (*posdf(bf(i))>posdf(v)*) **and**
      (*parent(bf(i))=0*) **and** (*posdf(bf(i))>posdf(bf(i-1))*) **and**
      (*bf(i)>bf(i-1)*)) **then**
      *parent(bf(i))=v*
      *last_son=bf(i); pos_last_son=i*
      **Compute**(*bf(i-1),posdf(bf(i))-1,posbf(df(posdf(bf(i-1))+1)),n*)
    **else break** // break the loop
  *i=posbf(df(posdf(last_son)+1))*
  **if** (*i>pos_last_son*) **then Compute**(*last_son, df_max_pos, i, n*)
**else return**

## 6 Related Work

We are not aware of any other attempts of inferring a company's structure based only on the number of pairs of employees having a certain property. Inferring hierachical structures based on several types of tree traversals has been achieved before, particularly for binary trees [2,3].

## 7 Conclusions and Future Work

In this paper we presented several algorithmic techniques for inferring the structure of a company (department structure, communication graph, hierarchy) using very limited available information. In most situations, the inferred structure is only one of the many possible structures which are consistent with the given input data, but, even so, it may be useful for performing a more insightful, in-depth analysis of the company.